\documentclass[12pt]{article}
\usepackage{graphicx}% Include figure files
\usepackage{bm}% bold math
\def\be{\begin{equation}}
\def\ee{\end{equation}}
\def\bea{\begin{eqnarray}}
\def\eea{\end{eqnarray}}
\usepackage{latexsym}
\usepackage{epsfig,amssymb,euscript}
\usepackage{amsmath}
\usepackage{array,calc,epsfig}
\newcommand{\Tr}{{\rm Tr}}
\usepackage{amsthm,amsfonts}
\usepackage{bbm}
\usepackage{graphicx}
\usepackage{braket}
\usepackage[makeroom]{cancel}
\usepackage{hyperref}
\begin{document}
\baselineskip=15.5pt
\pagestyle{plain}
\setcounter{page}{1}

\newfont{\namefont}{cmr10}
\newfont{\addfont}{cmti7 scaled 1440}
\newfont{\boldmathfont}{cmbx10}
\newfont{\headfontb}{cmbx10 scaled 1728}
\renewcommand{\theequation}{{\rm\thesection.\arabic{equation}}}
%\font\cmss=cmss10 \font\cmsss=cmss10 at 7pt

\vspace{1cm}

\begin{center}
{\huge{\bf Neutron-proton mass difference\\from gauge/gravity duality}}
\end{center}

\vskip 10pt

\begin{center}
{\large Francesco Bigazzi$^{a}$, Pierluigi Niro$^{b,c,d}$}\end{center}

\vskip 10pt
\begin{center}
\textit{$^a$ INFN, Sezione di Firenze; Via G. Sansone 1, I-50019 Sesto Fiorentino
(Firenze), Italy.}\\
\textit{$^b$ Dipartimento di Fisica e Astronomia, Universit\`a di
Firenze; Via G. Sansone 1, I-50019 Sesto Fiorentino
(Firenze), Italy.}\\
\textit{$^c$ Physique Th\'eorique et Math\'ematique and International Solvay Institutes,
Universit\'e Libre de Bruxelles; C.P. 231, 1050 Brussels, Belgium.}\\
\textit{$^d$ Theoretische Natuurkunde, Vrije Universiteit Brussel; Pleinlaan 2, 1050 Brussels, Belgium.}\\

\vspace{0.2cm}
{\small bigazzi@fi.infn.it, niro.pierluigi@gmail.com}
\end{center}

\vspace{25pt}

\begin{center}
 \textbf{Abstract}
\end{center}

\noindent 

Using gauge/gravity duality as a tool, we compute the strong sector, isospin breaking induced contribution to the neutron-proton mass difference in the Witten-Sakai-Sugimoto model of large $N$ QCD with two non-degenerate light flavors.  The mass difference, for which we provide an analytic expression, turns out to be positive and proportional to the down-up quark mass splitting, consistently with expectations and previous estimates based on effective QCD models. Extrapolating the model parameters to fit QCD hadronic observables, we find that the strong sector contribution to the nucleon mass splitting overcomes the electromagnetic contribution and is about $0.25\%$ of the average nucleon mass in the model, a result which approaches recent lattice QCD estimates. Our formula is extended to resonances and $\Delta$ baryons. We thus use it to compute the strong sector contribution to $\Delta$ baryons mass differences. Finally, we also provide details of how isospin breaking affects the holographic instanton solution describing the baryons.
\newpage

%%%%%%%%%%%%%%%%%%%%%%%%%%%%%%%%%%%%%%%
\section{Introduction}
\label{intro}
\setcounter{equation}{0}
We know very well from experiments \cite{PDG} that the neutron-proton mass splitting amounts to $M_n-M_p \approx 1.29 \text{ MeV}$. The sign and the magnitude of this difference, which is about $0.14\%$ of the average nucleon mass, are actually crucial features which have affected our Universe and its evolution in an essential way \cite{Wilczek}. 

If neutrons were lighter than protons the hydrogen atom would not be stable, due to proton decay. Moreover, a larger value of the mass splitting would have implied a very poor abundance of elements heavier than hydrogen, whose synthesis would have been more difficult. On the contrary, a smaller value would have implied that much more helium had been synthesized in the early Universe, leaving a poor quantity of hydrogen, which constitutes the fuel of nuclear fusion of stars. 

The neutron (resp. proton) constituents are one up and two down (resp. two up and one down) quarks. Even though there could be corrections arising from the (heavier) sea quarks, it is reasonable to expect that the isospin breaking mass splitting between the two constituent light quarks ($m_d-m_u\approx 2.5\,\rm{MeV}$ \cite{PDG}) is at the core of the strong sector contribution to the neutron-proton mass difference. The measured value of the difference is actually mostly\footnote{Other effects due to weak interactions or leptons provide negligible corrections.} the result of the competition between the latter and the electromagnetic contribution, which instead would lead the neutron to be lighter than the proton. In order to overcome the electromagnetic contribution, the strong interaction one has thus to be positive and larger in absolute value. 

The issue of an \textit{ab initio} computation of the neutron-proton mass difference is a challenging problem in hadron physics, due to the fact that it requires non-perturbative QCD tools to be used. The two main approaches adopted so far are based on chiral effective models and lattice techniques. 

The effective description of low-lying mesons and the analysis of the large $N$ limit of QCD led to the Skyrme picture of baryons as solitons of the chiral Lagrangian. A refinement of the original model based on just the pion effective action \cite{anw,an1} led to the introduction of the lightest (axial) vector mesons, $\rho$, $\omega$ (and $a_1$) to better account for near-core effects (see e.g. \cite{zahed} for a review). For instance, in the first and simplest of such extensions \cite{an2}, the introduction of the isosinglet meson $\omega$ was shown to stabilize the Skyrmions even without the addition of the higher-order Skyrme term. As it was shown in \cite{Jain}, the presence of vector mesons and the (two-flavor) $\eta$-type pseudoscalar excitation, induced at the quantum level, are the crucial ingredients to provide a non-vanishing strong contribution to the neutron-proton mass difference in a two-flavor model. The mass splitting contains a leading term which is linear in $m_d-m_u$ and turns out to be inversely proportional to the nucleon moment of inertia. As such it is a $1/N$ effect, as it happens to hyperfine baryon mass splittings \cite{anw}. Extrapolation of the model parameters to fit realistic hadronic observables gives $\Delta M_N^{\rm{strong}}\approx 1.3\, {\rm MeV}$ for the strong sector contribution to the nucleon mass splitting. The same parameters choice gives an average nucleon mass of the order of $1422\, {\rm MeV}$ \cite{Jain2}, so that the relative mass splitting in the model is about $0.09\%$.

Using improved QCD+QED lattice techniques, breakthrough progress on the computation of the nucleon mass splitting has been recently achieved in \cite{Borsanyi}. The computation takes into account the effects of four non-degenerate quarks (u, d, c, s) using a pion mass of about $195\,\rm{MeV}$. The strong-force contribution to the splitting, $\Delta M_N^{\rm{strong}}\approx 2.52\, {\rm MeV}$, which is about $0.27\%$ of the average nucleon mass, turns out to scale linearly with $m_d-m_u$ for small masses. The electromagnetic contribution is found to be $\Delta M_N^{\rm{e.m.}}\approx -1\,{\rm{MeV}}$.

In this work, we want to address the issue of the neutron-proton mass difference using a complementary approach based on the gauge/gravity holographic correspondence. The model we focus on, due to Witten \cite{witten}, Sakai and Sugimoto \cite{Sakai} (WSS) is, at low energies, the closest top-down relative to a large $N$ version of QCD ever built using string theory tools. In the large $N$ limit at strong coupling the model is described by a dual classical gravity background probed by $N_f$ $D8$-branes, where $N_f$ corresponds to the number of flavors. Unfortunately, when the classical gravity description is reliable, the dual field theory is coupled with spurious (Kaluza-Klein) matter fields in the adjoint representation and its UV behavior substantially deviates from that of real world QCD. Nevertheless, the gravity background is regular and given in closed analytic form and as such it allows to directly access (often with analytic control) highly non-trivial non-perturbative phenomena, like confinement, the emergence of a mass gap and chiral symmetry breaking, which are not spoiled by the above mentioned limitations. 

In the model, mesons correspond to open string fluctuations on the $D8$-branes. The lowest spin ones (pseudoscalar and vector mesons) are related to fluctuations of the $U(N_f)$ gauge field on the $D8$-branes. Baryons, instead, are interpreted as instanton configurations of the same gauge field \cite{Hata}, very much as the Skyrmions are solitons of the chiral Lagrangian. The notable feature of the model is that it automatically incorporates not only the pion chiral Lagrangian with the Skyrme term, but also the couplings with the whole tower of (axial) vector mesons. All the parameters in the mesonic effective action are given in terms of the few basic parameters of the model.

Here we consider the WSS model with $N_f=2$ non-degenerate light flavors, whose masses are accounted for by long open string instanton contributions as described in \cite{Aharony,hashimass}. To first order in the average quark mass and in the quark mass difference, we find an analytic expression for the (large $N$ leading contribution to the) strong part of the neutron-proton mass difference. This turns out to be positive, linear in $m_d - m_u$ and to scale as $1/N$ relatively to the isospin-preserving mass contribution to nucleon masses. Extrapolating the model parameters to fit real QCD hadronic data, we find that the mass splitting is about $0.25\%$ of the average nucleon mass in the model, a result which approaches the lattice estimate in \cite{Borsanyi}.\footnote{Just as it happens within the Skyrme model and in the original WSS setup \cite{Hata}, those extrapolations give less satisfactory quantitative results for what concerns the absolute value of the mass splitting (and of the average nucleon mass).} Our result can be complemented with a previous estimate \cite{Hong} of the electromagnetic contribution to the mass splitting in the WSS model, which turns out to be about $- 0.05\%$ of the average nucleon mass. Thus, consistently with expectations, in the WSS model the strong sector contribution to the neutron-proton mass difference overcomes the electromagnetic contribution.

The formula we find for the mass splitting is easily extended to other baryon states (and their excited levels) which, like the spin $3/2$ $\Delta$ ones, have only up and down quark constituents. We thus use our formula to estimate the relative mass splitting between charged and neutral $\Delta$ baryons as well.

Finally, although this is not necessary for the computation of the mass splittings, we provide details of how isospin breaking affects the holographic instanton solution describing the baryons.

The paper is organized as follows. In section \ref{rev} we briefly review the main ingredients of the holographic description of hadrons in the WSS model providing in turn a brief account of the effects induced by isospin breaking on the pseudoscalar meson mass spectrum and on the holographic instanton solution describing the baryons. Then, in section \ref{sic} we present our main result, the strong-force induced nucleon mass splitting. Alongside, we outline analogies and differences between the holographic approach and the Skyrme model one. Finally, we also consider the case of $\Delta$ baryons. We end up in section \ref{conc} with further comments on the obtained results and a comparison with those obtained in the WSS model with $N_f=3$ non-degenerate flavors in \cite{hashim3}. Further details are collected in the appendix.

\section{Holographic hadrons}
\label{rev}
\setcounter{equation}{0}
The WSS model is a four-dimensional $SU(N)$ gauge theory coupled to $N_f$ flavors of quarks and to massive adjoint Kaluza-Klein (KK) matter fields. The color degrees of freedom are the low energy modes of $N$ $D4$-branes wrapped on a circle of size $M_{KK}^{-1}$. The quarks arise from the low-lying modes of open strings stretching between the $D4$-branes and $N_f$ $D8$/$\bar{D}8$-branes. In the large $N$ limit, taking $N_f$ fixed, the strong coupling physics of the model is described by a dual regular classical gravity solution \cite{witten} probed by the $D8$-branes \cite{Sakai}. The backreaction of the latter can be actually neglected as a first approximation in the $N_f/N\ll1$ limit\footnote{See \cite{smearedWSS} for results going beyond this approximation.} and this amounts to treat the quarks in the so-called quenched regime. In the holographic model, highly non-trivial non-perturbative phenomena, like chiral symmetry breaking, confinement and the formation of a mass gap in the dual quantum field theory, can be easily (and often analytically) studied, providing new qualitative (and sometimes quantitative) insights on these IR features.

The WSS gravity solution also depends on a parameter $\lambda\sim T_s/M_{KK}^2$, where $T_s$ is the confining string tension. This parameter, which can also be thought of as the 't Hooft coupling at the scale $M_{KK}$, has to be very large in order for the classical gravity approximation to be reliable. This is actually one of the main limitations of the WSS model, since it implies that the interesting large $N$ QCD sector is not decoupled from the spurious massive Kaluza-Klein matter fields in that regime. 

The hadronic sector of the WSS model with massless quarks is holographically described by the effective action of the $D8$-branes \cite{Sakai}. Focusing just on QCD-like hadrons, the action can be reduced to the one which describes a  $U(N_f)$ Yang-Mills theory with Chern-Simons terms on a curved five-dimensional spacetime spanned by the 3+1 Minkowski coordinates $x^{\mu}$ and by the holographic radial direction $z\in(-\infty,\infty)$
\be\label{actions}
S_{WSS}= -\kappa\int d^4x d z\Tr\left(\frac{h(z)}{2} \mathcal{F}_{\mu\nu}\mathcal{F}^{\mu\nu} + k(z)\mathcal{F}_{\mu z}\mathcal{F}^\mu_{\;\; z}\right)+ \frac{N}{24\pi^2}\int  \Tr\left(\mathcal{A}\mathcal{F}^{ 2}- \frac{i}{2}\mathcal{A}^{ 3}\mathcal{F} - \frac{1}{10}\mathcal{A}^{ 5}\right)\,,
\ee
where the wedge product symbol ``$\wedge$" is understood. In the expression above, in units $M_{KK}=1$ 
(which we use from now on)
\be
\kappa = \frac{\lambda N}{216\pi^3}\,,\quad h(z) = (1+z^2)^{-1/3}\,,\quad k(z) = (1+z^2)\,.
\ee
The whole tower of massive vector (axial) mesons arises from the four-dimensional modes in a Kaluza-Klein reduction of the gauge field components ${\cal A}_{\mu}(x^{\mu},z)$. The $U(N_f)$ matrix, describing the pseudoscalar mesons, corresponds to the path-ordered holonomy matrix $\mathcal{U} = {\mathcal P} {\rm exp} [-i \int dz {\mathcal A}_z]$. Both the chiral Lagrangian (including the Skyrme and the WZW terms) and the Lagrangian for the whole (axial) vector meson tower are automatically contained in the simple action in (\ref{actions}). All the parameters in the effective theory can be expressed in terms of the few parameters of the WSS model. For instance the pion decay constant is given by
\be
f_{\pi}^2 = \frac{\lambda N}{54\pi^4}\,.
\ee 
Adding (small) quark mass terms to the model amounts to include (one) instanton corrections due to long open strings attached to the $D8$-branes \cite{Aharony,hashimass}.\footnote{An alternative proposal, based on the $D8$-${\bar D}8$ open string tachyon, can be found in \cite{sonnemass}.} As a result, the original action $S_{WSS}$ above can be completed by adding the mass term 
\be\label{actionmass}
S_M = c \int d^4x\, \Tr\mathcal{P}\left[M e^{-i\int \mathcal{A}_z dz}+ \mathrm{h.c.}\right]\,, \quad  c=\frac{\lambda^{3/2}}{3^{9/2}\pi^3}\,,
\ee
where $M$ is the quark mass matrix.\footnote{Here we choose for $c$ the normalization conventions adopted in \cite{hashimass}. See \cite{myersmass} for a detailed analysis of the mass term and related holographic renormalization issues.} Using the above mentioned map between the holonomy matrix $\mathcal {U}$ and the field ${\mathcal A}_z$, it is immediate to recover, in eq. (\ref{actionmass}), the same structure of the mass term $\Tr [M\mathcal{U}+\mathrm{h.c.}]$ usually introduced in the chiral Lagrangian approach. 

It is worth mentioning that the WSS model also accounts for a further term to be added to the effective action. It is induced by anomaly inflow and provides a Witten-Veneziano mass to the $\eta'$-like meson \cite{Sakai}. It reads
\be
S_{WV} = -\frac{\chi_g}{2}\int d^4x\left(\theta+\int dz \Tr\mathcal{A}_z\right)^2\,,\quad \chi_g=\frac{\lambda^3}{4(3\pi)^6}\,,
\label{seta}
\ee
where $\theta$ is the topological theta angle of the QFT (which can be rotated away by chiral rotations if massless quarks are present in the theory) and $\chi_g$ is the topological susceptibility of the unflavored theory. In the following we will take $\theta=0$ and a real and diagonal quark mass matrix. This implies that we are not considering topological effects and we assume that CP is conserved.\footnote{See \cite{wittentheta,theta1,theta2,theta3} for studies on $\theta$-dependent effects in the WSS model.}

Here we will focus on the two-flavor case $N_f=2$, decomposing the $U(2)$ gauge field as
\be
\mathcal A \equiv A + \frac{{\widehat A}}{2}\mathbf{1}\,,
\ee
where $A=A^a\tau^a/2$ is the $SU(2)$ component ($\tau^a$, $a=1,2,3$, being the Pauli matrices) and ${\widehat A}$ is the Abelian one. The two-flavor version of the $\eta'$ meson will be referred to as the $\eta$-type isosinglet in the following.

Since we are interested in the non-degenerate isospin breaking case, we will take the flavor mass matrix to be of the form
\begin{equation}
M=\begin{pmatrix}
m_u&0\\0&m_d
\end{pmatrix}=m \mathbbm{1}+\epsilon m\tau^3\,,
\label{mmatrix}
\end{equation}
where $m$ (which has to be assumed to be much smaller than $M_{KK}$) is the mean value of the quark masses
\begin{equation}
m=\frac{m_u+m_d}{2}\,,
\end{equation}
and $\epsilon$ parametrizes the deviation from the mass degenerate case
\begin{equation}
\epsilon=\frac{m_u-m_d}{m_u+m_d}\,.
\label{epseq}
\end{equation}
The $U(2)$ holonomy matrix $\mathcal{U}={\mathcal P}\exp\left(-i\int dz \mathcal{A}_z\right)$ can be decomposed as
\begin{equation}
\mathcal{U}=e^{i\varphi}U\,,
\end{equation}
where
\begin{equation}
\varphi=-\frac{1}{2}\int_{-\infty}^{+\infty}dz \widehat{A}_z\,,
\label{fi}
\end{equation}
is the $U(1)$ contribution, related to the $\eta$-type meson, and
\begin{equation}
U=\exp\left(-\frac{i}{2}\int_{-\infty}^{+\infty}dz A^a_z\tau^a \right)\,,
\label{ua}
\end{equation}
is the pure $SU(2)$ pion matrix, related to the pion triplet. 
\subsection{Mass splitting between charged and neutral pion}
\label{messplit}
Expanding $S_{WSS}+S_M+S_{WV}$ around the vacuum solution ${\mathcal A}=\mathbf 0$, both the Witten-Veneziano mass term for the $\eta$-type meson and the Gell-Mann-Oakes-Renner (GMOR) relation
\be
m_{WV}^2 = \frac{4\chi_g}{f_{\pi}^2}\,,\quad f_{\pi}^2 m_{\pi}^2 = 4 c m
\ee
are generated, to leading order in the small $m$ approximation. In the holographic regime $N,\lambda\gg1$, the validity of the quenched approximation and the small mass limit ($m\ll\Lambda_{QCD}$) imply that (in units $M_{KK}=1$) $m_{WV},m_{\pi}\ll 1$. Moreover, in order to reproduce the realistic case, the further limit $m_{\pi}\ll m_{WV}$ has to be taken. 

To leading order in $m$, the neutral and charged pions remain degenerate, despite the isospin breaking mass term. A (strong force induced) mass splitting can be recovered in the model working to subleading order in $m_{\pi}/m_{WV}$. The reason is the same as in the chiral Lagrangian case (see e.g. \cite{Jain}). If the isosinglet part of the matrix $\mathcal{U}$ is trivial in the vacuum, then the isospin breaking part of the mass term, proportional to $\epsilon\, Tr[\tau_3\mathcal{U}+\text{h.c.}]$ simply vanishes. However, the mass term induces a mixing between the $\eta$-type meson and the neutral pion: the isosinglet gets a non-vanishing ${\cal O}(\epsilon)$ v.e.v. and this, in turn, induces a ${\cal O}(\epsilon^2)$ mass splitting between charged and neutral pions. Let us review how does it works in detail.

In order to compute the pseudoscalar meson masses in the WSS model it is enough to expand $S_{WV}$ given in (\ref{seta}) and the mass term $S_M$ in (\ref{actionmass}) to quadratic order in the meson fields entering in 
\begin{equation}
\mathcal{U}=\exp\left[\frac{i}{f_{\pi}}\left(\pi^a\tau^a+S \mathbbm{1}\right)\right]\,,
\end{equation}
where we have taken the pion and the singlet decay constants to be equal, $f_{\pi}=f_{S}$, since we work in the large $N$ limit. All in all we get the following mass term in the Lagrangian 
\begin{equation}
\mathcal{L}_{(2)}=-\frac{1}{2}\frac{4cm}{f^2_{\pi}}\pi^a\pi^a-\frac{1}{2}\left(\frac{4cm}{f^2_{\pi}}+m^2_{WV}\right)S^2-\frac{4c\epsilon m}{f^2_{\pi}} S \pi^3\,.
\end{equation}
We thus see that the mass matrix is diagonal only for the charged pions. Thus, the actual neutral pion $\pi^0$ and the $\eta$-type meson are given by the linear combinations of $\pi^3$ and $S$ which diagonalize the mass matrix. This yields the following mass spectrum (to leading order in $m^2_{\pi}/m^2_{WV}\ll1$)
\begin{equation}
\begin{split}
&m^2_{\pi^{\pm}}=m^2_{\pi}\,, \\
&m^2_{\pi^0}=m^2_{\pi}\left(1-\epsilon^2 \frac{m^2_{\pi}}{m^2_{WV}}\right)\,, \\
&m^2_{\eta}=m^2_{WV}+m^2_{\pi}\left(1+\epsilon^2 \frac{m^2_{\pi}}{m^2_{WV}}\right)\,.
\end{split}
\label{mesonmasses}
\end{equation}
Thus, including the $U(1)$ singlet in the holonomy matrix explicitly gives a strong force induced splitting between the masses of the neutral and the charged pion
\begin{equation}
m^2_{\pi^{\pm}}-m^2_{\pi^0}=\epsilon^2 \left(\frac{m^2_{\pi}}{m^2_{WV}}\right) m^2_{\pi}\,.
\end{equation}
This mass splitting is proportional to the square of the isospin breaking parameter, as it should be since the quark composition of the pions, and hence their mass difference, is invariant under the exchange of the two quarks (or under $\epsilon\rightarrow -\epsilon$). 
  %%%%%%%%%%%%%%%%%%%%
\subsection{Holographic baryons}
Taking inspiration from the Skyrme model, and more precisely motivated by the stringy identification \cite{wittenbaryon} of baryon vertices with wrapped $D4$-branes, in the WSS model baryons arise as heavy instanton solutions of the effective five-dimensional action \cite{Hata}. The instanton number
\be
n_B=\frac{1}{8\pi^2}\int_B \Tr \mathcal{F}\wedge\mathcal{F}\,,
\ee
where $B$ is the space spanned by $x^{M}\equiv (x^{1,2,3},\,z)$, is then interpreted as the baryon number. 

The one instanton solution in the $N_f=2$ WSS model with massless flavors is analytically known around $z=0$; the general solution is presented in \cite{Hashimoto}. In the former limit, the static solution is just a BPST instanton \cite{BPSTInst} which, due to the presence of the Chern-Simons terms, also sources an electric potential. In the ``singular gauge" it reads
\bea \label{instanton}
&&A_M^{\mathrm{cl}} = -i (1-f(\xi)) g^{-1} \partial_M g\,, \qquad A_0^{\mathrm{cl}}=\widehat{A}^{\mathrm{cl}}_M=0\,,\nonumber \\
&&\widehat{A}^{\mathrm{cl}}_0 = \frac{N}{8\pi^2\kappa}\frac{1}{\xi^2}\left[1-\frac{\rho^4}{(\rho^2+\xi^2)^2}\right]\,,
\eea
where $\xi^2 \equiv (z-Z)^2 + |\vec{x}-\vec{X}|^2$ and 
\begin{equation}
f(\xi) = \frac{\xi^2}{\xi^2+\rho^2}\,, \quad g(x) = \frac{(z-Z)\mathbf 1 - i (\vec{x}-\vec{X})\cdot \vec{\tau}}{\xi}\,.
\end{equation}
The solution depends on various parameters: the coordinates of the center of mass of the baryon in the usual three-dimensional space $\vec{X}$ and in the holographic fifth direction $Z$, the size of the instanton $\rho$ and, more implicitly, the set $\lbrace a_I\rbrace$ ($I=1,2,3,4$) of parameters which parametrize global $SU(2)$ transformations of the instanton solution
\begin{equation}
\mathbf{a}=a_4\mathbbm{1}+i\sum_{i=1}^3a_i\tau^i, \quad\quad \sum_{I=1}^4 a^2_I = 1\,.
\end{equation}

It is worth outlining the parallel between the static instanton solution in WSS and the soliton solution in the Skyrme model with low-lying vector mesons. In \cite{Jain}, for instance, the static soliton solution is described by the pion matrix $U$, the space components of the isotriplet vector ($\rho_i$) and by the time component of the isosinglet vector ($\omega_0$): these correspond, respectively, to the (lowest modes of the) fields $A_z$, $A_i$, $\widehat A_0$ which span the BPST-like solution (\ref{instanton}).

Substituting the solution (\ref{instanton}) into the WSS original action (\ref{actions}) one finds $S_{WSS,\,\rm{on\,shell}} = -\int d t M_B$, where, up to $\mathcal{O}(\lambda^{-2})$ corrections \cite{Hata}
\begin{equation}\label{MB}
M_B(\rho,Z) =  M_0\left[1 +  \left(\frac{\rho^2}{6}+ \frac{N^2}{320 \pi^4 \kappa^2}\frac{1}{\rho^2} + \frac{Z^2}{3}\right)\right]\,,\quad M_0\equiv8\pi^2 \kappa\,,
\end{equation}
with $M_0$ giving the baryon mass in the $N\rightarrow\infty$, $\lambda\rightarrow\infty$ limit. 

This tells us that while $\vec{X}$ and the gauge group orientations are genuine instanton moduli, $\rho$ and $Z$ are not. Minimization of  (\ref{MB}) with respect to these parameters gives their classical values
\begin{equation}
\rho_{cl}^2=\frac{N}{8\pi^2\kappa}\sqrt{\frac{6}{5}}=\frac{27\pi}{\lambda}\sqrt{\frac{6}{5}}, \quad\quad Z_{cl}=0\,,
\label{classicalmoduli}
\end{equation}
and the leading order result
\begin{equation}
M_{B\,\text{cl}}=8\pi^2\kappa \left(1+\frac{1}{\lambda}\frac{54\pi}{\sqrt{30}}\right)\,.
\label{mclassic}
\end{equation}
The above relations imply that the center of the instanton is classically localized at $Z=0$ and its size $\rho \sim 1/\sqrt{\lambda}$ is very small but finite in the $\lambda\gg1$ regime. A crucial role in stabilizing the baryon size is played by the Coulomb term $\widehat A_0^{\rm{cl}}$ precisely as it happens in the Skyrme model with the (time component of the) isosinglet vector meson $\omega$ \cite{an2}.

The quantum states of the baryons are obtained as eigenstates of the corresponding quantum Hamiltonian, in which the parameters of the instanton solution are promoted to be time-dependent operators in a non-relativistic Quantum Mechanics. The non-relativistic approximation is justified by the fact that the baryon mass $M_B\sim N\lambda$ is parametrically large with respect to quantum fluctuations. The quantization, just as in the Skyrme model \cite{anw}, acts by rotating the static classical solution for the $SU(2)$ pion matrix as
\begin{equation}
U_{\rm{cl}}\rightarrow\mathbf{a}(t)U_{\rm{cl}}\mathbf{a}(t)^{\dagger}\,,
\label{isorotation}
\end{equation}
where, for a classical instanton localized at $\vec{X}=0$, using (\ref{ua}) and (\ref{instanton}) we have
\be
U_{\rm{cl}}\equiv u_0\mathbbm{1}-iu_i\tau^i =-\cos\alpha\mathbbm{1}-i\sin\alpha\frac{x^i}{r}\tau^i\,,
\label{umatrix}
\end{equation}
where, in turn,
\begin{equation}
\alpha=\alpha(r)=\frac{\pi}{\sqrt{1+\rho^2/r^2}}\,,
\label{alphadef}
\end{equation}
 with $r=|\vec x|$.\footnote{The above result, obtained using the BPST-like instanton solution (which is strictly valid around $z\approx 0$) is a very good approximation of the result obtained by using the complete instanton solution (as found in \cite{Hashimoto}). The difference between the two is in fact negligible at large $N$ and $\lambda$. For our purposes it will be enough to extrapolate the small-$z$ results to the whole $z$ axis.}
 
Time-dependent rotations analogous to the one above act non-trivially also on the isotriplet $A_i$ and trivially on the isosinglet $\widehat A_0$. Moreover, for consistency with the equations of motion, novel components of the gauge field are generated which depend on the time derivatives of the instanton parameters. Thus, in the complete time-dependent instanton solution, also the isotriplet $A_0$ and the isosinglet $\widehat A_i$ and $\widehat A_z$ components are turned on \cite{Hashimoto}. This is again precisely analogous to what happens in the Skyrme model with vector mesons examined in \cite{Jain}: there the time-dependent soliton solution also includes non-trivial expressions for the time component of the isotriplet ($\rho_0$), the space components of the isosinglet vector ($\omega_i$) and the $\eta$-type meson. The contribution of the latter, which turns out to be proportional to the scalar product $\vec\chi\cdot\vec x$, where $\vec\chi$ is the angular velocity of the spinning soliton, is actually crucial to provide a mass splitting between neutron and proton in \cite{Jain}. As it will be evident in the following, the same is true for the WSS model. 

In detail, in the $z\ll1$ region, we have, for the time dependent non-abelian field components (in the singular gauge)
\bea
A_0 &=& -i (1-f(\xi)) \mathbf{a}\,\mathbf{\dot a}^{-1} + i(1-f(\xi)) {\dot X}^M \mathbf{a}(g^{-1}\partial_M g)\mathbf{a}^{-1}\,,\nonumber \\
A_M &=& -i(1-f(\xi)) \mathbf{a}(g^{-1}\partial_M g) \mathbf{a}^{-1}\,.
\label{eqnonab}
\eea
The velocity dependent abelian field components read (again, let us consider the  $z\approx0$ region)
\begin{equation}
\widehat{A}_z=-\frac{N}{8\pi^2\kappa} \left[ \frac{\xi^2+2\rho^2}{(\xi^2+\rho^2)^2}\dot{Z} + \frac{\rho^2}{(\xi^2+\rho^2)^2} \left(\frac{\chi^j x^j}{2}+\frac{\dot{\rho}z}{\rho}\right) \right]\,,
\label{hataz}
\end{equation}
and
\begin{equation}
\widehat{A}_i=-\frac{N}{8\pi^2\kappa} \left[ \frac{\xi^2+2\rho^2}{(\xi^2+\rho^2)^2}\dot{X}^i + \frac{\rho^2}{(\xi^2+\rho^2)^2} \left(\frac{\chi^a}{2}(\epsilon^{iaj} x^j-\delta^{ia}z)+\frac{\dot{\rho}x^i}{\rho}\right) \right]\,,
\label{hatai}
\end{equation}

Notice that eq. (\ref{hataz}) precisely contains a term $\vec\chi\cdot\vec x$ where
\begin{equation}
\chi^j=-i\mathrm{Tr}(\tau^j\mathbf{a}^{-1}\mathbf{\dot{a}})=\frac{J^{j}}{4\pi^2\kappa\rho^2}\,,
\label{chij}
\end{equation}
is the angular velocity and $J^j$ is the spin operator 
\be
J_k = \frac{i}{2} \left( - y_4 \frac{\partial}{\partial y_k} + y_k \frac{\partial}{\partial y_4} - \epsilon_{klm} y_l \frac{\partial}{\partial y_m}\right)\,,
\label{spin}
\ee
where $y_I\equiv\rho a_I$ and $k,l,m=1,2,3$. The quantum Hamiltonian is obtained substituting the time-dependent solution into the original WSS action. The isospin operator 
\begin{equation}
I_k = \frac{i}{2} \left( + y_4 \frac{\partial}{\partial y_k} - y_k \frac{\partial}{\partial y_4} - \epsilon_{klm} y_l \frac{\partial}{\partial y_m}\right)\,,
\label{isospin}
\end{equation}
can also be rewritten as
\be
- i \Tr (\tau^a \mathbf{a}\,\mathbf{\dot{a}}^{-1}) = \frac{I^a}{4\pi^2\kappa\rho^2}\,,
\ee
and it is such that only states with $I=J$ appear in the spectrum. 
In order to quantize the instantons as fermions an anti-periodicity condition $\psi(a^I) = -\psi(-a^I)$ has to be implemented on the baryon wave functions. The related states have $I=J=\ell/2$ with $\ell=1,3,5,\cdots$ positive odd integers. 

A baryon state $|B, s\rangle$ will depend on the the (iso)spin and on further quantum numbers $n_\rho$ and $n_Z$ which describe excited states or resonances; the case $\ell=1$, $n_\rho = n_Z = 0$ corresponds to the unexcited neutron (with isospin component $I_3=-1/2$) and proton ($I_3=1/2$) and the corresponding wave functions are
\bea
&& |p\uparrow\rangle \propto R(\rho) \psi_Z(Z) (a_1 + i a_2)\,,\quad  |p\downarrow\rangle \propto R(\rho) \psi_Z(Z) (a_4 - i a_3)\,,\nonumber\\
&& |n\uparrow\rangle\propto R(\rho) \psi_Z(Z) (a_4+ i a_3)\,,\quad  |n\downarrow\rangle \propto R(\rho) \psi_Z(Z)(a_1 - i a_2)\,,
\label{pnwave}
\eea
with
\be
R(\rho)=\rho^{-1+2\sqrt{1+N_c^2/5}}e^{-\frac{M_0}{\sqrt{6}}\rho^2} \, ,\qquad \psi_Z(Z)=e^{-\frac{M_0}{\sqrt{6}}Z^2} \,.
\end{equation}
Generalizations to larger values of $\ell, n_{\rho}, n_{Z}$ can be found in \cite{Hata}.

The above results hold at zero-th order in the quark mass. To first order in the small $m$ limit we can treat the quark mass term $S_{M}$ in (\ref{actionmass}) as a perturbation and compute the related effects on the instanton solution and the baryon Hamiltonian. In the isospin-preserving case the leading order effect on the baryon mass has already been computed in \cite{Hirayama2}. Here we will see what happens including isospin breaking terms. 

In the following subsection we will provide details of how the instanton solution describing WSS baryons is modified by $\epsilon$-dependent corrections, where $\epsilon$ is the isospin breaking parameter in (\ref{epseq}). Then, in section \ref{sic}, we will compute the $\cal{O}({\epsilon})$ isospin breaking term in the baryon Hamiltonian. In this case it will be enough to compute the on-shell value of the $\epsilon$-dependent term in the action $S_M$ using the unperturbed ($\epsilon$-independent) instanton solution presented above. Moreover, to first order in $\epsilon$, the corresponding Hamiltonian eigenvalue will be obtained (according to standard perturbation theory) by computing the matrix element of the isospin breaking Hamiltonian term on the unperturbed nucleon state.
%%%%%%%%%%%%%%%%%
\subsection{Isospin breaking effects on holographic baryons}
Let us now show how the static instanton configuration describing baryons is modified by the presence of the isospin breaking mass term. The equations of motion which follow from the action $S_{WSS}+S_M+S_{WV}$ read
\begin{equation}
-\kappa\left[h(z)\partial_{\nu}\widehat{F}^{\mu\nu}+\partial_z\left(k(z)\widehat{F}^{\mu z}\right)\right]+\frac{N}{128\pi^2}\epsilon^{\mu\alpha\beta\gamma\delta}\left(F^a_{\alpha\beta}F^a_{\gamma\delta}+\widehat{F}_{\alpha\beta}\widehat{F}_{\gamma\delta}\right)=0\,,
\label{1}
\end{equation}
\begin{equation}
-\kappa\left[h(z)D_{\nu}F^{\mu\nu}+D_z\left(k(z)F^{\mu z}\right)\right]^a+\frac{N}{64\pi^2}\epsilon^{\mu\alpha\beta\gamma\delta}\widehat{F}_{\alpha\beta}F^a_{\gamma\delta}=0\,,
\label{2}
\end{equation}
\bea
&&-\kappa k(z)\partial_{\mu}\widehat{F}^{z \mu}+\frac{N}{128\pi^2}\epsilon^{z\mu\nu\rho\sigma}\left(F^a_{\mu\nu}F^a_{\rho\sigma}+\widehat{F}_{\mu\nu}\widehat{F}_{\rho\sigma}\right)=\nonumber \\
&& =\chi_g \int dz\widehat{A}_z+ic\mathrm{Tr}\mathcal{P}\left(\frac{M}{2}\mathcal{U}-\text{h.c.}\right)\,,
\label{3}
\eea
\begin{equation}
-\kappa k(z)\left(D_{\mu}F^{z \mu}\right)^a+\frac{N}{64\pi^2}\epsilon^{z\mu\nu\rho\sigma}\widehat{F}_{\mu\nu}F^a_{\rho\sigma}=ic\mathrm{Tr}\mathcal{P}\left(M\frac{\tau^a}{2}\mathcal{U}-\text{h.c.}\right)\,.
\label{4}
\end{equation}
We want to solve these equations perturbatively in the mass correction. Thus, we plug in
\begin{equation}
\mathcal{F}=\mathcal{F}^{(0)}+\mathcal{F}^{(mass)},
\end{equation}
where $\mathcal{F}^{(0)}$ is the instanton solution in the absence of mass term, which satisfies the equations of motion at zeroth order, while $\mathcal{F}^{(mass)}$ is the mass perturbation which is of $\mathcal{O}(m)$. 

Focusing again in the near-flat region $z \ll 1$, the relevant equations to solve are (at leading order in the large $N$ and $\lambda$ expansion)
\begin{equation}
-\kappa\left[h(z)\partial_{\nu}\widehat{F}^{0\nu}_{mass}+\partial_z\left(k(z)\widehat{F}^{0z}_{mass}\right)\right]-\frac{N}{32\pi^2}\epsilon^{ijk}\left(F^a_{ij}F^{mass,a}_{kz}+F^a_{iz}F^{mass,a}_{jk}\right)=0,
\label{abeliantime}
\end{equation}
\begin{equation}
-\kappa\left[h(z)\partial_{\nu}\widehat{F}^{i\nu}_{mass}+\partial_z\left(k(z)\widehat{F}^{iz}_{mass}\right)\right]=0,
\label{abelianspace}
\end{equation}
\begin{equation}
-\kappa k(z)\partial_{\nu}\widehat{F}^{z\nu}_{mass}-\chi_g\int{dz\widehat{A}^{mass}_z}=2cm\epsilon \frac{x^3}{r}\sin\alpha,
\label{abelianz}
\end{equation}
\begin{equation}
-\kappa\left[h(z)D_{\nu}F^{0\nu}+D_z\left(k(z)F^{0z}\right)\right]^a\big|_{mass}-\frac{N}{32\pi^2}\epsilon^{ijk}\left(F^a_{ij}\widehat{F}^{mass}_{kz}+F^a_{iz}\widehat{F}^{mass}_{jk}\right)=0,
\label{nonabeliantime}
\end{equation}
\begin{equation}
-\kappa\left[h(z)D_{\nu}F^{i\nu}+D_z\left(k(z)F^{iz}\right)\right]^a\big|_{mass}=0,
\label{nonabelianspace}
\end{equation}
\begin{equation}
-\kappa k(z)\left(D_{\nu}F^{z\nu}\right)^a\big|_{mass}=2cm\frac{x^a}{r}\sin\alpha.
\label{nonabelianz}
\end{equation}
where $\alpha$ is given in (\ref{alphadef}). The notation $\big|_{mass}$ means to take only the linear contribution in $m$, while for sake of simplicity we have denoted the unperturbed solution without the subscript $(0)$. We are interested in static solutions, which means that $\mathcal{A}^{mass}$ for the moment has no time dependence, and hence $\nu$ indices will be only spatial indices\footnote{This is trivially true for standard derivatives $\partial_{\mu}$. For covariant derivatives, in ($\ref{nonabelianspace}$) and ($\ref{nonabelianz}$) we have a term like $D_{\mu}F^{M\mu}=D_jF^{Mj}+i[A_0+A^{mass}_0,F^{M0}+F^{M0}_{mass}]$. The second term vanishes at linear order in $m$ because both $A_0$ and $F^{M0}$ vanish on the unperturbed static solution.}. 
%%%%%%%%%%%%%%%%%%%%%%%%%%%%%%%%%%%%%%%%%%%%%
\subsubsection{Solution for $\widehat{A}_i^{mass}$ and $\widehat{A}_z^{mass}$}
It is easy to realize that a consistent solution for the space component $\widehat{A}_i^{mass}$ is given by
\begin{equation}
\widehat{A}_i^{mass}=0\,.
\label{null}
\end{equation} 
Using the above result we get that eq. ($\ref{abelianz}$) becomes
\begin{equation}
\kappa k(z)\partial_i\partial_i\widehat{A}_z^{mass}-\chi_g\int{dz\widehat{A}^{mass}_z}=  - c(m_d-m_u) \frac{x^3}{r}\sin\alpha\,,
\label{azmass}
\end{equation} 
whose solution can be found in the form
\begin{equation}
\widehat{A}_z^{mass}=-\frac{c (m_d-m_u)}{\kappa} \frac{b(r)}{k(z)}\frac{x^3}{r}\,,
\end{equation} 
Notice that
\begin{equation}
\frac{\chi_g}{\kappa}\int_{-\infty}^{+\infty}\frac{dz}{k(z)}=m^2_{WV}\,,
\end{equation}
so that the second term of the left-hand side of ($\ref{azmass}$) would produce a contribution which is subleading in our approximations and hence we can neglect it. Plugging the ansatz for $\widehat{A}_z^{mass}$ into ($\ref{azmass}$) we get the following equation for $b(r)$
\begin{equation}
\frac{1}{r^2}\left[\partial_r(r^2\partial_r b(r))-2b(r)\right]=\sin\alpha\,.
\label{b}
\end{equation}
The solution is
\begin{equation}
\begin{split}
b(r)=&\int_{0}^{+\infty}dr'b_G(r,r')\sin\frac{\pi}{\sqrt{1+\rho^2/r'^2}}\,,\\
&b_G(r,r')=\begin{cases}
-\frac{r}{3} &\mathrm{if}\;r'>r\,,\\
-\frac{r'^3}{3r^2} &\mathrm{if}\;r'<r\,.
\end{cases}
\end{split}
\label{bsolution}
\end{equation}
It is easy to deduce how this solution is modified in the time-dependent case, taking into account how eq. (\ref{isorotation}) enters in the source term for $\widehat{A}_z^{mass}$. We find 
\begin{equation}
\widehat{A}_z^{mass}=-\frac{c (m_d-m_u)}{\kappa} \frac{b(r)}{k(z)}\frac{\Tr[\tau^3\mathbf{a}\tau^i\mathbf{a}^{-1}]x^i}{2 r}\,,
\end{equation}
which scales as 
\begin{equation}
\widehat{A}_z^{mass}\sim \frac{f_{\pi}^2 m_{\pi}^2}{N\lambda^2}\frac{m_d-m_u}{m_u+m_d}\,,
\end{equation}
in the $z\ll 1$ region.
%%%%%%%%%%%%%%%%%%%%%%%%%%%%%%%%%%%%%
\subsubsection{An overview on the other gauge field components}
Let us now consider equation ($\ref{nonabeliantime}$) for the non-abelian component $A_0^{mass}$. It reads
\begin{equation}
-\left[h(z)D_{\nu}F^{0\nu}+D_z\left(k(z)F^{0z}\right)\right]\big|_{mass}=\frac{N}{32\pi^2\kappa}\epsilon^{ijk}\left(F_{ij}\hat{F}^{mass}_{kz}+F_{iz}\hat{F}^{mass}_{jk}\right).
\end{equation}
We do not try to solve this equation at the moment, but we want to extrapolate the scaling of the solution in the $z\ll 1$ region. The left-hand side scales as the second spatial derivative of $A_0^{mass}$, while the right-hand side, which acts as a source term, reads
\begin{equation}
RHS=- \frac{27\pi}{\lambda}\frac{c (m_d-m_u)}{\kappa}\frac{\rho^2}{(\xi^2+\rho^2)^2}\frac{1}{k(z)}\vec{\tau}\cdot\vec{\nabla}\left(b(r)\frac{x^3}{r}\right)\,.
\end{equation}
This implies that, in the $z\ll1$ region, $A_0^{mass}$ scales as
\be
A_0^{mass}\sim \frac{f_{\pi}^2m_{\pi}^2}{N\lambda^{5/2}}\frac{m_d-m_u}{m_d+m_u}\,.
\ee
Equations ($\ref{nonabelianspace}$) and ($\ref{nonabelianz}$) are coupled differential equations for $A_i^{mass,a}$ and $A_z^{mass,a}$, in which the instanton solution and mass term act as sources. However, there is no $\epsilon$ dependence, so the discussion of the qualitative behavior of the solutions follows as in the mass degenerate case, see \cite{theta3}. We will not discuss these equations further.

Finally, equation ($\ref{abeliantime}$) reads
\begin{equation}
h(z)\partial_i\partial_i\widehat{A}^0_{mass}+\partial_z(k(z)\partial_z\widehat{A}^0_{mass})=\frac{27\pi}{4\lambda}\epsilon^{ijk}\left(F^a_{ij}F^{mass,a}_{kz}+F^a_{iz}F^{mass,a}_{jk}\right),
\end{equation}
from which we obtain the solution for $\widehat{A}^0_{mass}$ after having solved the equations for the spatial components of the non-abelian field $A_i^{mass,a}$ and $A_z^{mass,a}$, which act as source terms. Thus, also this component will not receive corrections in $m_d-m_u$. 
%%%%%%%%%%%%%%%%%%%%%%%%%%%%%%%%%%%%%%%%%
\section{Strong interaction contribution to the nucleon mass splitting}
\label{sic}
\setcounter{equation}{0}
Let us now come to the core of this paper and compute the isospin-breaking term in the baryon Hamiltonian, to leading order in $m$ and in the $\lambda\gg1$, $N\gg1$ limit. To first order in $\epsilon$, this term can be simply deduced from the on-shell value (on the unperturbed massless instanton solution, see in particular eq. (\ref{eqnonab}) and (\ref{hataz})) of the isospin breaking term in the Lagrangian   
(see eq. (\ref{actionmass}))
\be
L^{i.b.} = c\,m\,\epsilon \int d^3x\,\mathrm{Tr}(\tau^3e^{-\frac{i}{2}\int dz {\widehat A}_z}\mathbf{a}U_{\text{cl}}\mathbf{a}^{\dagger}+\text{h.c.})\,.
\label{Ls}
\ee
It is easy to realize that in the static case, where $\widehat A_z=0$, the on-shell value of this term is zero. The velocity dependent component $\widehat A_z$ (see eq. (\ref{hataz})) plays a crucial role in producing a non-trivial result, similarly to what happens for the velocity dependent $\eta$-type meson in the Skyrme model computation in \cite{Jain}. As $\widehat A_z\sim1/(N\sqrt{\lambda})$, we will work in the adiabatic limit and approximate the related exponential in (\ref{Ls}) as $e^{-(i/2)\int dz {\widehat A}_z}\approx 1 -(i/2)\int dz {\widehat A}_z$. 

To leading order in our approximations, the induced isospin breaking term in the baryon Hamiltonian, which we will write as $H=H_0+H^{i.b.}$, where $H_0$ contains all the isospin-preserving terms,  turn out to scale as 
\be
H^{i.b.} \sim \frac{f_{\pi}^2 m_{\pi}^2}{N\lambda^{5/2}}\frac{m_d-m_u}{m_d+m_u} I_3\,,
\label{scalH}
\ee  
To first order in the above coefficient, the nucleon mass difference will thus be obtained as
\begin{equation}
M_n-M_p=\bra{n}H^{i.b.}\ket{n}-\bra{p}H^{i.b.}\ket{p}\,,
\end{equation}
where $\ket{n}$ and $\ket{p}$ are the unperturbed nucleon states (\ref{pnwave}).

In the adiabatic limit, the on-shell value of (\ref{Ls}) to first order in $\epsilon$ is given by
\be
L^{i.b.} = -\frac{cN(m_d-m_u)}{48\kappa}\rho^3\, \Tr[\tau^3\mathbf{a}\tau^i\mathbf{a}^{-1}]\chi^i {\cal J}_2\,,
\label{Lisob}
\ee
where
\be
\mathcal{J}_2=\int_{0}^{+\infty}dy\frac{1}{(1+y^{-2})^{3/2}}\sin\left(\frac{\pi}{\sqrt{1+y^{-2}}}\right)\approx1.054\,.
\ee
The isospin breaking Lagrangian term (\ref{Lisob}) is linear in the angular velocities $\chi^i$ and as such it modifies the canonically conjugate momentum, i.e. the angular momentum $J^i$, so that\footnote{We thank Stefano Bolognesi and Lorenzo Bartolini for relevant related comments.}
\be
J^i = 4\pi^2 \kappa \rho^2 \chi^i  - \frac{cN(m_d-m_u)}{48\kappa}\rho^3\, \Tr[\tau^3\mathbf{a}\tau^i\mathbf{a}^{-1}]{\cal J}_2\equiv 4\pi^2 \kappa \rho^2 \chi^i - K^i\,.
\ee
Correspondingly, the angular momentum term in the Hamiltonian is modified as
\begin{equation}
H_J = \frac12 \frac{(\vec J +\vec K)^2}{4\pi^2\kappa\rho^2}\,,
\end{equation}
and the isospin breaking term is easily read as
\begin{equation}
H^{i.b.}= \frac{\vec J\cdot\vec K}{4\pi^2\kappa\rho^2}= - \frac{cN(m_d-m_u)}{96\pi^2\kappa^2}\rho\, {\cal J}_2 (I_3\otimes\mathbbm{1})\,,
\label{quamasham}
\end{equation}
where we have used the well known relation (whose derivation we review in appendix \ref{app1}) between spin and isospin operators 
\begin{equation}
J^i\mathrm{Tr}[\tau^a\mathbf{a}\tau^i\mathbf{a}^{-1}]=-2I^a\,,
\end{equation}
and we have explicitly put the identity operator when required. In the tensor product in (\ref{quamasham}) the operator on the left acts on isospin space, while the operator on the right acts on spin space. In this way we have found a term which is linear in $m_d-m_u$ and proportional to the third component of the isospin operator. In the WSS model, at large $N$ and $\lambda$ and to first order in $m$ and in the isospin breaking parameter $\epsilon$, this gives the {\it leading} contribution to the mass difference between baryon states with the same spin $J=\ell/2$ which would form isospin multiplets if not for the isospin breaking term. Neglecting for a moment the issue of $\rho$ quantization, it reads (reinserting the dependence on $M_{KK}$)
\begin{equation}
\Delta M_B^{\rm{strong}}= -\Delta I_3 \frac{f_{\pi}^2m_{\pi}^2}{N\lambda^{5/2}M_{KK}^3}\,\frac{m_d-m_u}{m_d+m_u}\,729\,\pi^{9/2}\left(\frac{54}{5}\right)^{1/4} {\cal J}_2\,.
%\frac{18\pi^{3/2}}{N\lambda}\left(\frac{6}{5}\right)^{1/4}\mathcal{J}_2(m_d-m_u)\,.
\label{gencla}
\end{equation}
In the case of the neutron-proton mass difference $\Delta I_3 =-1$
\be
M_n-M_p=\frac{f_{\pi}^2m_{\pi}^2}{N\lambda^{5/2}M_{KK}^3}\,\frac{m_d-m_u}{m_d+m_u}\,729\,\pi^{9/2}\left(\frac{54}{5}\right)^{1/4} {\cal J}_2\,.
%\frac{18\pi^{3/2}}{N\lambda}\left(\frac{6}{5}\right)^{1/4}\mathcal{J}_2(m_d-m_u)\,.
\label{npcla}
\end{equation}
Hence, the neutron-proton mass difference is found to be positive and linear in $m_d-m_u$, as expected. Moreover, as anticipated in (\ref{scalH}), it is $1/N$ suppressed with respect to the isospin preserving nucleon mass term \cite{Hirayama2}, as it happens in the Skyrme approach where it depends on the inverse of the nucleon moment of inertia \cite{Jain}.

Although for large $N$ and $\lambda$ the results above (with $\rho=\rho_{\rm{cl}}$) are certainly enough to capture the leading order behavior, in principle one should remember that $\rho$ is a quantum operator. Thus, we should replace its classical value by its expectation value on the quantum baryon states. The difference between the two is subleading at large $N$, but once we extrapolate the model to $N=3$ this is not true anymore (see e.g. \cite{Hashimoto}). The correct quantum version of eq. (\ref{gencla}) is thus (for states with the same $\ell, n_{\rho}, n_{Z}$ quantum numbers)
\begin{equation}
\Delta M_B^{\rm{strong}}(\ell, n_{\rho},\Delta I_3)=-\Delta I_3\frac{c N (m_d-m_u)}{96\pi^2\kappa^2}\mathcal{J}_2\langle\rho\rangle_{\ell\,,n_{\rho}}\,.\label{genquant}
\end{equation}
At the same time, taking into account the quantization of the pseudomoduli $\rho$ and $Z$,  the WSS average mass of baryons with spin-isospin $J=I=\ell/2$, to first order in the average quark mass $m$, up to ${\cal O}(\epsilon_f)$ corrections due to flavor backreaction ($\epsilon_f\sim\lambda^2 N_f/N\ll1$ in the probe approximation), neglecting electromagnetic contributions, reads\footnote{Here we also neglect possible terms arising from zero-point energy contributions, as discussed in \cite{Hata}. We thank Shigeki Sugimoto for pointing out this issue.} 
%(reinserting the $M_{KK}$ dependence)
\cite{Hata,Hirayama2}
\bea
\overline{M}_B(\ell,n_{\rho},n_{Z}) &=&  \left(8\pi^2\kappa+\sqrt{\frac{(\ell+1)^2}{6}+\frac{2}{15}N^2}+\sqrt{\frac23}(n_{\rho}+n_Z+1)\right)M_{KK}+\nonumber\\
&&+16\pi c\, m {\cal J}_1\langle\rho^3\rangle_{\ell\,,n_{\rho}}\,,
\label{mavq}
\eea
where the last term is the shift due to the average quark mass $m$ \cite{Hirayama2} and
\be
\mathcal{J}_1=\int_{0}^{\infty}{dy y^2\left[\cos\left(\frac{\pi}{\sqrt{1+y^{-2}}}\right)+1\right]}\approx1.104\,.
\ee
The explicit form of eq. (\ref{genquant}) and (\ref{mavq}) requires knowledge of the expectation value of (powers of) $\rho$ on the related baryon state (see e.g. \cite{hashim3}). For $n_{\rho}=0$,
\be
\langle\rho^n\rangle_{\ell\,,n_{\rho}=0} = \frac{\Gamma[\beta_{\ell} + n/2]}{c_{\rho}^{n/2}\Gamma[\beta_{\ell}]}\,,
\ee
where 
\be
\beta_{\ell} \equiv 1+\sqrt{(\ell+1)^2 +\frac{4N^2}{5}}\,,\quad c_{\rho} \equiv \frac{16\pi^2\kappa}{\sqrt{6}}\,.
\ee
For $n_{\rho}=1$,
\be
\langle\rho^n\rangle_{\ell\,,n_{\rho}=1} =\left(1+\frac{n(n+2)}{4\beta_{\ell}}\right)\langle\rho^n\rangle_{\ell\,,n_{\rho}=0}\,.
\ee
For larger values of $n_{\rho}$ the related expressions can be easily obtained using e.g. the results collected in section 2 of \cite{hashim3}.

In order to compare our results with real QCD, it is worth considering dimensionless quantities. Let us focus on the ground states $n_{\rho}=n_{Z}=0$ and choose the following ratio
\begin{equation}
\gamma(\ell,\Delta I_3)\equiv\frac{\Delta M_B^{\rm strong}(\ell,0,\Delta I_3)}{\overline{M}_B(\ell,0,0)}\,.
\end{equation}
Let us now extrapolate the above results to $N=3$, choosing for the remaining parameters the values most commonly used in the literature to fit WSS expressions with realistic mesonic observables\footnote{Just as it happens in the Skyrme models, see e.g. \cite{Jain2}, the WSS absolute baryon masses, computed using the above parameters, are notoriously larger (e.g. about 2 times for the nucleons) than the actual experimental values. This is one of the reasons why we find more interesting to focus on relative mass values.} $f_{\pi}=92.4$ MeV and $m_{\rho}=776$ MeV:
\begin{equation}
\lambda=16.63, \quad M_{KK}=949\text{ MeV}\, .
\end{equation}
We also fix the quark masses as
\begin{equation}
m=3.14\text{ MeV}, \quad m_d-m_u = 2.5\text{ MeV}\, ,
\end{equation}
where $m$ reproduces the charged pion mass $m_{\pi^{\pm}}=140$ MeV using the relation (\ref{mesonmasses}) and $m_d-m_u$ is taken from the current experimental estimate in \cite{PDG}. From these we get, for the relative strong force contribution to the neutron-proton mass splitting
\be
\gamma(1, -1)\equiv 2\frac{M_n-M_p}{M_n+M_p}\approx 0.25\,\%\,.
\label{npqu}
\ee
Notice that using the leading order expression (\ref{npcla}) we would get $M_n-M_p\approx 5.54\,{\rm MeV}$ which is about $0.33\%$ the average classical nucleon mass
\be
\overline{M}_N^{\rm{cl}}=N\left(\frac{\lambda}{27\pi}+\frac{2}{\sqrt{30}}\right)M_{KK} + 4\pi\, \frac{f_{\pi}^2m_{\pi}^2}{M_{KK}^3}\,{\cal J}_1\left(\frac{27\pi}{\lambda}\right)^{3/2}\left(\frac65\right)^{3/4}\,.
%\left(\frac65\right)^{3/4}\frac{16}{\sqrt{\pi}}{\cal J}_1\,m\,.
\ee
Our result (\ref{npqu}) can be compared with the value obtained using the measured masses of the neutron ($939.56$ MeV) and the proton ($938.27$ MeV) and the recent lattice result \cite{Borsanyi}, which is $\Delta M_N^{\text{strong}}=2.52$ MeV.\footnote{See also \cite{sh}.} This gives
\begin{equation}
\gamma(1, -1)_{\text{QCD}}\approx 0.27\,\%\,,
\end{equation}
which approaches our result. 

Moreover we get 
\be
\gamma(3, -2)\equiv 2\frac{M_{\Delta^{0}}-M_{\Delta^{++}}}{M_{\Delta^{0}}+M_{\Delta^{++}}}\approx 0.48\,\%\,,
\ee
for the relative strong force contribution to the $\Delta^{0}$-$\Delta^{++}$ mass splitting (on which we focus since it can be estimated experimentally \cite{PDG}). In turn we get that 
\be
\frac{M_{\Delta^{0}}-M_{\Delta^{++}}}{M_n-M_p}\approx 2.33\,,
\ee
to be compared with the experimental data \cite{PDG} giving about $2.6/1.29\approx 2.02$ for the same ratio and with the leading order result, obtained fixing $\rho=\rho_{\rm{cl}}$, giving exactly 2.
%%%%%%%%%%%%%%%%%
\section{Comments}\label{conc}
\setcounter{equation}{0}
Our result is in the framework of a large $N$ QCD model with $N_f=2$ light flavors. In real QCD, considering only the two valence light quarks is expected to be a reasonable approximation for the computation of the strong sector contribution to the neutron-proton (or $\Delta$ baryons) mass difference. Contributions from heavier quarks dynamics are expected to provide minor corrections. However, in large $N$ QCD (both in effective chiral theories \cite{Jain} and in the holographic WSS model investigated in this work) the two-flavor result emerges as a quantum effect suppressed in $1/N$, due to the full time-dependent baryon configuration. On the contrary, the $N_f=3$ result emerges already at the level of the classical (i.e. static) solution. For the WSS model it reads \cite{hashim3}
\be
\Delta M_B^{\rm{strong}}(l,n_{\rho},(p,q), a_k)=a_k\frac{8\pi c (m_d-m_u)}{3}\mathcal{J}_1\langle\rho^3\rangle_{l,n_{\rho},(p,q)}\,,
\label{3fla}
\ee
where, for $n_{\rho}=0$, $a_k=1/5$ (resp. $1/2$) for the neutron-proton (resp. $\Delta^0$-$\Delta^{++}$) mass difference,
\be
\langle\rho^3\rangle_{l,n_{\rho}=0, (p,q)}=\frac{\Gamma[\beta_{l,(p,q)}+3/2]}{c_{\rho}^{3/2}\Gamma[\beta_{l,(p,q)}]}\,,
\ee
where $(p,q)=(1,1), l=1$ for the nucleons and $(p,q)=(3,0), l=3$ for the $\Delta$ baryons and
\be
\beta_{l,(p,q)}=1+\sqrt{\frac{49}{4}+\frac{2N^2}{15}+\frac83(p^2+q^2+3(p+q)+pq)-l(l+2)}\,.
\ee
The expression (\ref{3fla}) does not depend on the strange quark mass, thus we cannot extrapolate from it a two-flavor result like (\ref{genquant}) in the $m_s \gg m_u,m_d$ limit. Indeed, also the $N_f=3$ results are obtained in the small mass limit, so that formally taking $m_s\rightarrow\infty$ is not allowed in any case. 

In the large $N$ limit, the three-flavor result would seem to dominate with respect to the two-flavor one. However, there are various concerns about this point. First, the $N_f=3$ results could be quantitatively modified by the inclusion of higher order corrections in the strange quark mass. Moreover, in the usual extrapolation to $N=3$, using the above relations and the results in \cite{hatamurata,hashim3} to get the $N_f=3$ expression for the WSS baryon mass, one finds that the relative nucleon mass splitting is about $0.085\%$ of the average nucleon mass. This is much smaller than the $N_f=2$ WSS and the lattice results. A similar feature is also present in the Skyrme-like models, see e.g.  section IV in \cite{Jain}. This seems to confirm the expectation that the main contribution to the nucleon (and $\Delta$) mass splitting should be captured in an $N_f=2$ model.
%%%%%%%%%%%%%%%%%%%%%%%%%%%%%%%%%%%%%%%%%%%%%%%%
\vskip 15pt \centerline{\bf Acknowledgments} \vskip 10pt \noindent We are deeply indebted to Stefano Bolognesi for his contributions in the early stage of the project, for having pointed out to our attention the paper \cite{Jain} and for many relevant discussions and comments. We are grateful to Aldo L. Cotrone for many relevant comments and discussions and for a careful reading of a preliminary version of this work. We thank Lorenzo Bartolini for a careful check on some of the results presented in this work and for useful comments. We thank Shigeki Sugimoto for clarifying comments on the baryon masses in the WSS model. The work of P.N. is financially supported by a scholarship of the International Solvay Institutes. 
\appendix
\section{Relation between spin and isospin operators}
\label{app1}
\setcounter{equation}{0}
We want to prove explicitly that
\begin{equation}
\mathrm{Tr}(\tau^a\mathbf{a}\tau^i\mathbf{a}^{-1})J_i=-2I_a\,,
\end{equation}
i.e. that
\begin{equation}
\frac{1}{\rho^2}\mathrm{Tr}(\tau^a\mathbf{y}\tau^i\mathbf{y}^{-1})J_i=-2I_a\,.
\label{result}
\end{equation}
The spin and isospin operators are given by
\begin{equation}
\begin{split}
J_i &= \frac{i}{2} \left( - y_4 \partial_i + y_i \partial_4 - \epsilon_{ilm} y_l \partial_m \right)\,, \\
I_a &= \frac{i}{2} \left( + y_4 \partial_a - y_a \partial_4 - \epsilon_{abc} y_b \partial_c\right)\,,
\end{split}
\end{equation}
where $\partial_i\equiv\partial/\partial y_i$. Moreover, $\mathbf{y}$ is a $SU(2)$ matrix which can be decomposed as
\begin{equation}
\mathbf{y}=y_4\mathbbm{1}+iy_k\tau^k\,, \quad\quad y^2_4+y^2_k = \rho^2\,,
\end{equation}
so that
\begin{equation}
\mathbf{y}^{-1}=y_4\mathbbm{1}-iy_s\tau^s\,.
\end{equation}
Here the sum over $k,s=1,2,3$ is understood. The last ingredient we need to prove the relation (\ref{result}) is given by the trace of Pauli matrices
\begin{equation}
\begin{split}
\mathrm{Tr}(\tau^a\tau^i)&=2\delta^{ai}\,,\\
\mathrm{Tr}(\tau^a\tau^i\tau^k)&=2i\epsilon^{aik}\,,\\
\mathrm{Tr}(\tau^a\tau^k\tau^i\tau^s)&=2(\delta^{ak}\delta^{is}+\delta^{as}\delta^{ik}-\delta^{ai}\delta^{ks})\,.
\end{split}
\end{equation} 
We are now ready to prove (\ref{result}): first we compute
\begin{equation}
\begin{split}
\frac{1}{\rho^2}\mathrm{Tr}(\tau^a\mathbf{y}\tau^i\mathbf{y}^{-1})
&=\frac{1}{\rho^2}\mathrm{Tr}\left[(y_4\tau^a+iy_k\tau^a\tau^k)(y_4\tau^i-iy_s\tau^i\tau^s)\right]=\\
&=\frac{1}{\rho^2}\left[y^2_4\mathrm{Tr}(\tau^a\tau^i)-2iy_4y_k\mathrm{Tr}(\tau^a\tau^i\tau^k)+y_ky_s\mathrm{Tr}(\tau^a\tau^k\tau^i\tau^s)\right]=\\
&=\frac{2}{\rho^2}\left[(y^2_4-y_k^2)\delta_{ai}+2y_ay_i+2\epsilon_{aik}y_4y_k\right]\,.
\end{split}
\end{equation}
Then, we apply this to the spin operator to obtain
\begin{equation}
\begin{split}
\frac{1}{\rho^2}\mathrm{Tr}(\tau^a\mathbf{y}\tau^i\mathbf{y}^{-1})J_i
&=\frac{2}{\rho^2}\frac{i}{2}\left[(y^2_4-y_k^2)\delta_{ai}+2y_ay_i+2\epsilon_{aik}y_4y_k\right]\times \\
&\times\left(-y_4\partial_i+y_i\partial_4-\epsilon_{ilm}y_l\partial_m\right)=\\
&=\frac{2}{\rho^2}\frac{i}{2}\left[y_a(y^2_4-y^2_k+2y_k^2)\partial_4-y_4(y^2_4-y^2_k+2y^2_k)\partial_a+\right.\\
&\left.-\epsilon_{alm}y_l(y^2_4-y^2_k)\partial_m-2\epsilon_{aml}y_ly^2_4\partial_m\right]=\\
&=-2\times\frac{i}{2}(y_4\partial_a-y_a\partial_4-\epsilon_{alm}y_l\partial_m)=-2I_a\,,
\end{split}
\end{equation}
which proves our initial claim. We have also verified that $[J_i, \mathrm{Tr}(\tau^a\mathbf{a}\tau^i\mathbf{a}^{-1})]=0$ so that 
\be
J_i\mathrm{Tr}(\tau^a\mathbf{a}\tau^i\mathbf{a}^{-1})=-2I_a\,.
\ee
%%%%%%%%%%%%%%%%%%%%%%%%%%%%%%%%%%%%%%%%%%%%%%%%%%%%%%%%%%%%%%%%%%%%%%%


\begin{thebibliography}{99}
\bibitem{PDG}C.~Patrignani {\it et al.} [Particle Data Group],
  ``Review of Particle Physics,''
  Chin.\ Phys.\ C {\bf 40}, no. 10, 100001 (2016).
% doi:10.1088/1674-1137/40/10/100001
  %%CITATION = doi:10.1088/1674-1137/40/10/100001;%%
\bibitem{Wilczek} F.~Wilczek,
  ``Particle physics: A weighty mass difference,''
  Nature {\bf 520}, 303 (2015).
%  doi:10.1038/nature14381
  %%CITATION = doi:10.1038/nature14381;%%
\bibitem{anw}G.~S.~Adkins, C.~R.~Nappi and E.~Witten,
  ``Static Properties of Nucleons in the Skyrme Model,''
  Nucl.\ Phys.\ B {\bf 228}, 552 (1983).
%  doi:10.1016/0550-3213(83)90559-X
  %%CITATION = doi:10.1016/0550-3213(83)90559-X;%%
 \bibitem{an1} G.~S.~Adkins and C.~R.~Nappi,
  ``The Skyrme Model with Pion Masses,''
  Nucl.\ Phys.\ B {\bf 233}, 109 (1984).
 % doi:10.1016/0550-3213(84)90172-X
  %%CITATION = doi:10.1016/0550-3213(84)90172-X;%%
\bibitem{zahed} I.~Zahed and G.~E.~Brown,
 ``The Skyrme Model,''
  Phys.\ Rept.\  {\bf 142}, 1 (1986).
 % doi:10.1016/0370-1573(86)90142-0
  %%CITATION = doi:10.1016/0370-1573(86)90142-0;%%  
\bibitem{an2}G.~S.~Adkins and C.~R.~Nappi,
  ``Stabilization of Chiral Solitons via Vector Mesons,''
  Phys.\ Lett.\  {\bf 137B}, 251 (1984).
%  doi:10.1016/0370-2693(84)90239-9
  %%CITATION = doi:10.1016/0370-2693(84)90239-9;%%
\bibitem{Jain}P.~Jain, R.~Johnson, N.~W.~Park, J.~Schechter and H.~Weigel,
  ``The Neutron - Proton Mass Splitting Puzzle in Skyrme and Chiral Quark Models,''Phys.\ Rev.\ D {\bf 40}, 855 (1989).
%  doi:10.1103/PhysRevD.40.855
  %%CITATION = doi:10.1103/PhysRevD.40.855;%%
\bibitem{Jain2} P.~Jain, R.~Johnson, Ulf-G.~Meissner, N.~W.~Park and J.~Schechter,
 ``Realistic Pseudoscalar Vector Chiral Lagrangian and Its Soliton Excitations,''
  Phys.\ Rev.\ D {\bf 37}, 3252 (1988).
 % doi:10.1103/PhysRevD.37.3252
  %%CITATION = doi:10.1103/PhysRevD.37.3252;%% 
\bibitem{Borsanyi}S.~Borsanyi {\it et al.},
  ``Ab initio calculation of the neutron-proton mass difference,''
  Science {\bf 347}, 1452 (2015)
%  doi:10.1126/science.1257050
  [arXiv:1406.4088 [hep-lat]].
  %%CITATION = doi:10.1126/science.1257050;%%
\bibitem{witten}E.~Witten,
  ``Anti-de Sitter space, thermal phase transition, and confinement in gauge theories,''
  Adv.\ Theor.\ Math.\ Phys.\  {\bf 2}, 505 (1998)
%  doi:10.4310/ATMP.1998.v2.n3.a3
  [hep-th/9803131].
  %%CITATION = doi:10.4310/ATMP.1998.v2.n3.a3;%%  
\bibitem{Sakai}T.~Sakai and S.~Sugimoto,
  ``Low energy hadron physics in holographic QCD,''
  Prog.\ Theor.\ Phys.\  {\bf 113}, 843 (2005)
 % doi:10.1143/PTP.113.843
  [hep-th/0412141].
  %%CITATION = doi:10.1143/PTP.113.843;%%
 \bibitem{Hata}H.~Hata, T.~Sakai, S.~Sugimoto and S.~Yamato,
  ``Baryons from instantons in holographic QCD,''
  Prog.\ Theor.\ Phys.\  {\bf 117}, 1157 (2007)
 % doi:10.1143/PTP.117.1157
  [hep-th/0701280 [HEP-TH]].
  %%CITATION = doi:10.1143/PTP.117.1157;%%
\bibitem{Aharony}O.~Aharony and D.~Kutasov,
  ``Holographic Duals of Long Open Strings,''
  Phys.\ Rev.\ D {\bf 78}, 026005 (2008)
 % doi:10.1103/PhysRevD.78.026005
  [arXiv:0803.3547 [hep-th]].
  %%CITATION = doi:10.1103/PhysRevD.78.026005;%%  
 \bibitem{hashimass}K.~Hashimoto, T.~Hirayama, F.~L.~Lin and H.~U.~Yee,
 ``Quark Mass Deformation of Holographic Massless QCD,''
  JHEP {\bf 0807}, 089 (2008)
 % doi:10.1088/1126-6708/2008/07/089
  [arXiv:0803.4192 [hep-th]].
  %%CITATION = doi:10.1088/1126-6708/2008/07/089;%% 
\bibitem{Hong}  D.~K.~Hong,
  ``Holographic Estimate of Electromagnetic Mass,''
  JHEP {\bf 1508}, 066 (2015)
 % doi:10.1007/JHEP08(2015)066
  [arXiv:1409.8139 [hep-ph]].
  %%CITATION = doi:10.1007/JHEP08(2015)066;%% 
\bibitem{hashim3}
K.~Hashimoto, N.~Iizuka, T.~Ishii and D.~Kadoh,
  ``Three-flavor quark mass dependence of baryon spectra in holographic QCD,''
  Phys.\ Lett.\ B {\bf 691}, 65 (2010)
 % doi:10.1016/j.physletb.2010.06.008
  [arXiv:0910.1179 [hep-th]].
  %%CITATION = doi:10.1016/j.physletb.2010.06.008;%%
\bibitem{smearedWSS}F.~Bigazzi and A.~L.~Cotrone,
  ``Holographic QCD with Dynamical Flavors,''
  JHEP {\bf 1501}, 104 (2015)
%  doi:10.1007/JHEP01(2015)104
  [arXiv:1410.2443 [hep-th]].
  %%CITATION = doi:10.1007/JHEP01(2015)104;%%
\bibitem{sonnemass}O.~Bergman, S.~Seki and J.~Sonnenschein,
  ``Quark mass and condensate in HQCD,''
  JHEP {\bf 0712}, 037 (2007)
 % doi:10.1088/1126-6708/2007/12/037
  [arXiv:0708.2839 [hep-th]].
\bibitem{myersmass}R.~McNees, R.~C.~Myers and A.~Sinha,
  ``On quark masses in holographic QCD,''
  JHEP {\bf 0811}, 056 (2008)
 %doi:10.1088/1126-6708/2008/11/056
  [arXiv:0807.5127 [hep-th]].
\bibitem{wittentheta}E.~Witten,
``Theta dependence in the large N limit of four-dimensional gauge theories,''
Phys.\ Rev.\ Lett.\  {\bf 81}, 2862 (1998)
%doi:10.1103/PhysRevLett.81.2862
[hep-th/9807109].
 %%CITATION = doi:10.1103/PhysRevLett.81.2862;%%  
\bibitem{theta1}F.~Bigazzi, A.~L.~Cotrone and R.~Sisca,
 ``Notes on Theta Dependence in Holographic Yang-Mills,''
  JHEP {\bf 1508}, 090 (2015)
 % doi:10.1007/JHEP08(2015)090
  [arXiv:1506.03826 [hep-th]].
  %%CITATION = doi:10.1007/JHEP08(2015)090;%%
\bibitem{theta2}L.~Bartolini, F.~Bigazzi, S.~Bolognesi, A.~L.~Cotrone and A.~Manenti,
 ``Neutron electric dipole moment from gauge/string duality,''
  Phys.\ Rev.\ Lett.\  {\bf 118}, no. 9, 091601 (2017)
 % doi:10.1103/PhysRevLett.118.091601
  [arXiv:1609.09513 [hep-ph]].
  %%CITATION = doi:10.1103/PhysRevLett.118.091601;%%
\bibitem{theta3} L.~Bartolini, F.~Bigazzi, S.~Bolognesi, A.~L.~Cotrone and A.~Manenti,
  ``Theta dependence in Holographic QCD,''
  JHEP {\bf 1702}, 029 (2017)
 % doi:10.1007/JHEP02(2017)029
  [arXiv:1611.00048 [hep-th]].
  %%CITATION = doi:10.1007/JHEP02(2017)029;%%
  \bibitem{wittenbaryon}E.~Witten,
 ``Baryons and branes in anti-de Sitter space,''
  JHEP {\bf 9807}, 006 (1998)
%  doi:10.1088/1126-6708/1998/07/006
  [hep-th/9805112].
  %%CITATION = doi:10.1088/1126-6708/1998/07/006;%%
\bibitem{Hashimoto} K.~Hashimoto, T.~Sakai and S.~Sugimoto,
  ``Holographic Baryons: Static Properties and Form Factors from Gauge/String Duality,''
  Prog.\ Theor.\ Phys.\  {\bf 120}, 1093 (2008)
 % doi:10.1143/PTP.120.1093
  [arXiv:0806.3122 [hep-th]].
  %%CITATION = doi:10.1143/PTP.120.1093;%%
\bibitem{BPSTInst} A.~A.~Belavin, A.~M.~Polyakov, A.~S.~Schwartz and Y.~S.~Tyupkin,
 ``Pseudoparticle Solutions of the Yang-Mills Equations,''
  Phys.\ Lett.\ B {\bf 59}, 85 (1975).
%  doi:10.1016/0370-2693(75)90163-X
  %%CITATION = doi:10.1016/0370-2693(75)90163-X;%%  
\bibitem{Hirayama2}K.~Hashimoto, T.~Hirayama and D.~K.~Hong,
  ``Quark Mass Dependence of Hadron Spectrum in Holographic QCD,''
  Phys.\ Rev.\ D {\bf 81}, 045016 (2010)
%  doi:10.1103/PhysRevD.81.045016
  [arXiv:0906.0402 [hep-th]].
  %%CITATION = doi:10.1103/PhysRevD.81.045016;%%
\bibitem{sh} R.~Horsley {\it et al.},
 ``Isospin splittings of meson and baryon masses from three-flavor lattice QCD + QED,''
  J.\ Phys.\ G {\bf 43}, no. 10, 10LT02 (2016)
%  doi:10.1088/0954-3899/43/10/10LT02
  [arXiv:1508.06401 [hep-lat]].
  %%CITATION = doi:10.1088/0954-3899/43/10/10LT02;%%
\bibitem{hatamurata}H.~Hata and M.~Murata,
 ``Baryons and the Chern-Simons term in holographic QCD with three flavors,''
  Prog.\ Theor.\ Phys.\  {\bf 119}, 461 (2008)
%  doi:10.1143/PTP.119.461
  [arXiv:0710.2579 [hep-th]].
  %%CITATION = doi:10.1143/PTP.119.461;%%
  %16 citations counted in INSPIRE as of 23 Feb 2018
\end{thebibliography}
\end{document}